\documentclass[aps,preprint,a4paper,showpacs,amsmath,amssymb,floatfix]{revtex4}
\usepackage{graphicx}
\usepackage{dcolumn}
\usepackage{bm}
\newcommand{\Op}[1]{{\boldsymbol{\mathrm{\hat{#1}}}}}
\newcommand{\beq}{\begin{equation}}
\newcommand{\eeq}{\end{equation}}
\newcommand{\beqar}{\begin{eqnarray}}
\newcommand{\eeqar}{\end{eqnarray}}
\newcommand{\bea}{\begin{eqnarray}}
\newcommand{\eea}{\end{eqnarray}}
\newcommand{\bcen}{\begin{center}}
\newcommand{\ecen}{\end{center}}

\begin{document}
\draft

\title{Optimal Performance of  Reciprocating Quantum Refrigerators }

\author{Tova Feldmann and Ronnie Kosloff}

\address{
Institute  of Chemistry
the Hebrew University, Jerusalem 91904, Israel\\
}

\begin{abstract}
A reciprocating quantum refrigerator is studied with the purpose
of determining the limitations of cooling to absolute zero. We find that if the energy spectrum 
of the working medium possesses an uncontrollable gap, then there is a minimum achievable temperature above zero.
Such a gap, combined with a negligible amount of noise, prevents adiabatic following
during the demagnetization stage which is the necessary condition for reaching $T_c \rightarrow 0$. 
The refrigerator is based on an  Otto cycle where the working medium is an interacting spin system 
with an energy gap. For this system the external control Hamiltonian does not commute with the internal interaction.
As a result during the demagnetization and magnetization segments of the operating cycle the system cannot follow adiabatically the temporal change in the energy levels. We connect the nonadiabatic dynamics to quantum friction.
An adiabatic measure is defined characterizing the rate
of change of the Hamiltonian.
Closed form solutions are found for a constant adiabatic measure for all the cycle segments.
We have identified  a family of quantized frictionless cycles with increasing cycle times. These cycles minimize the entropy production. 
Such frictionless cycles are able to cool to $T_c=0$. External noise on the controls eliminates these frictionless cycles.
The influence of phase and amplitude noise on the demagnetization and magnetization segments is explicitly derived.
An extensive numerical study of optimal cooling cycles was carried out which showed that at sufficiently low temperature
the noise always dominated restricting the minimum temperature.
\end{abstract}
\maketitle
\section{Introduction} 
\label{sec:introduction}

Reciprocating refrigerators operate by a working medium shuttling heat from the cold to the hot reservoir. The task is carried out by a controlled dynamical system. A change in the Hamiltonian of the system is accompanied by a change in the internal temperature.
Upon contact with the cold side the temperature of the working medium  is forced to be
lower than $T_c$-the cold bath temperature.  A reciprocal relation is required on the hot side.
Explicitly a quantum refrigerator is studied where the control of temperature is 
governed by manipulating the energy levels of the system. 

One of the main characterization of a refrigerator is the minimum temperature it can reach.
The third law of thermodynamics already restricts this temperature to be the absolute zero
\cite{nerst06b,nerst18}. Practically the minimum temperature is determined by the details of the mechanism of the  heat pump. 
To investigate the cooling  problem we study a model 
of a reciprocating quantum refrigerator. The present study is a comprehensive account following 
a brief version \cite{tova09b}.
The main issues to be addressed are:
\begin{itemize}
\item{What are the restrictions imposed by the working medium?}
\item{What are the optimal conditions required to reach the minimum temperature?}
\item{Is there a minimum temperature above the absolute zero?}
\end{itemize}

To gain insight on these issues a reverse Otto cycle is considered where the working medium consists of interacting spin system. The magnetization/demagnetization stages are  carried out by varying an external magnetic field which alters the energy levels of the working medium.
Such a model is a simplified version of adiabatic demagnetization refrigerator (ADR) \cite{oja97,kurti82,hakonen91}. 
These refrigerators have found use in cooling detectors to very low temperatures in space missions but also in an attempt to replace the existing technology in home appliances
\cite{lchen06,gschneider05,rowea06}. In addition there is a growing interest in 
quantum engines and refrigerators \cite{geusic67,k24,k122,k156,k169,lloyd,he02,bender,kieu04,segal06,bushev06,erez07,erez08,mahler08,jahnkemahler08,allahmahler08,segal09,he09} 
with the purpose of unraveling the relation between quantum mechanics and thermodynamics.
The present paper follows a series of studies on a first principle four stroke 
quantum engines \cite{k85,k87,k116,k152,k176,k190,k201,k215,k221}, where
it was demonstrated that the model engines displays the irreversible 
characteristics of common engines operating in finite time \cite{salamon01}. 

A generic working medium possesses a Hamiltonian that is only partially controlled externally:
\begin{equation}
{\Op H }~~=~~{\Op H}_{int} + {\Op H}_{ext}(\omega)
\label{eq:genhamil}
\end{equation} 
where $\omega=\omega(t)$ is the time dependent external control field. 
Typically,  $[{\Op H}_{int},{\Op H}_{ext}] \ne 0$., therefore, $[\Op H(t), \Op H(t')] \ne 0$ as a result 
a state diagonal in the temporary energy eigenstates cannot follow adiabatically the changes due to the control. 
The inability of the state to follow the change in the energy spectrum is source of quantum friction
 \cite{k176,k190,k201,k215,k221}. This friction limits  the 
performance of the heat engine as well as the heat pump.  
There is an intimate connection between adiabatic following and the ability to reach cold temperatures. 
Since friction limits the performance almost perfect adiabaticity is the key to low temperature refrigeration.
In this study we will explore the prospects of almost frictionless refrigeration cycles.
Typically, the internal interaction in the working medium  leads to an uncontrollable  finite gap $J$
in the energy level spectrum between the ground and first excited state.  
We will show that this gap combined with unavoidable quantum friction will be linked to a finite minimal temperature.

A good characterization of the deviation from adiabticity is the difference between the von Neumann entropy of the state ${\cal S}_{vn}=-tr \{\Op \rho \log \Op \rho \}$ and the energy entropy defined by the projections on the energy eigenstate ${\cal S}_E=-\sum{ p_j \log p_j}$, where $p_j$ is the population of energy state $j$. 
Equality is obtained only for perfect adiabatic following  and thermal equilibrium \cite{k190}.

The present study explores the properties of quantum first principle 
four stroke heat pumps, with emphasis on the  approach to absolute zero.  
In a previous study based on a  phenomenological heat pump \cite{k152}
we have found a linear relation between the cooling rate and the cold bath temperature $T_c$.
Does this relation survive when first principle quantum dynamical consideration
are accounted for?

\section{The Cycle of Operation, the Quantum Heat Pump}
\label{sec:cycle}

The working medium in the present study is composed of an interacting spin system.
Eq. (\ref{eq:genhamil}) is modeled by the ${\bf SU}(2)$ algebra of operators.  We can realize the model by a system of two coupled spins
where the internal interaction is described by:
\begin{equation}
{\Op H}_{int} ~~=~~\frac {1} {2} \hbar J \left({ {\boldsymbol{\mathrm{\hat
{\sigma}}}}_x^1} \otimes { {\boldsymbol{\mathrm{\hat{\sigma}}}}_x^2} -
{{\boldsymbol{\mathrm{\hat{\sigma}}}}_y^1}\otimes {\boldsymbol
{\mathrm{\hat{\sigma}}}}_y^2 ~~~
  \right) ~\equiv~\hbar J {\Op B_2}
  \label{eq:onter}
 \end{equation}
where ${{\boldsymbol{\mathrm{\hat{\sigma}}}}}$ represents the spin-Pauli 
operators, and $J$ scales the strength of the inter particle interaction. 
For $J \rightarrow 0$, the system approaches 
a working medium with noninteracting atoms \cite{k152}.
The external Hamiltonian represents interaction of spins with an external magnetic field:
\begin{equation}
{\Op H}_{ext} ~~=~~\frac {1}{2}\hbar \omega(t)
\left({\boldsymbol{\mathrm{\hat{\sigma}}}}_z^1
\otimes {\bf \hat I^2}
+
{\bf \hat I^1} \otimes {{{{\boldsymbol{ \mathrm {  \sigma}}}}}_z^2}
\right)~\equiv~\omega(t) {\Op B_1}~~.
\label{eq:ext}
\end{equation}
The ${\bf SU(2)}$ is closed with ${\Op B_3}
~~=~~\frac{1}{2}  \left({ {\boldsymbol{\mathrm{\hat
{\sigma}}}}_y^1} \otimes { {\boldsymbol{\mathrm{\hat{\sigma}}}}_x^2} +
{{\boldsymbol{\mathrm{\hat{\sigma}}}}_x^1}\otimes {\boldsymbol
{\mathrm{\hat{\sigma}}}}_y^2 ~~~
  \right)$ and $[ {\Op B_1}, {\Op B_2}] \equiv  2 i {\Op B_3}$.
  
The total Hamiltonian modeling Eq. (\ref{eq:genhamil}) then becomes: 
\begin{equation}
{\Op H}=\hbar \left(\omega(t) {\bf \hat B_{1}}+\rm J {\bf \hat B_{2}}\right)~~.
\label{eq:hamil}
\end{equation}
The adiabatic energy levels, the eigenvalues of $\Op H(t)$ are $ \epsilon_1= - \hbar {\Omega} ,~
\epsilon_{2/3}=0,~ \epsilon_4= \hbar {\Omega} $ where $\Omega=\sqrt{\omega^2+J^2}$. For $J \ne 0$
there is a zero field splitting, an irreduceable gap between the ground and excited state levels. Eq. (\ref{eq:hamil}) contains the essential features of the Hamiltonian
of magnetic materials \cite{oja97}.

The dynamics of the quantum thermodynamical observables are described by 
completely positive maps within the formulation of quantum open systems
\cite{lindblad76,alicki87,breuer} . The dynamics is generated by the Liouville 
superoperator, $ {\cal L}$, studied in the Heisenberg picture,
\begin{equation}
\frac {d {\Op A}}{dt}~~=~~ \frac{i}{\hbar}[{\Op H}, {\Op A}]+ {\cal L}_{D}( {\Op A})
~+~ \frac{\partial {\Op A}}{\partial t}~~~.
\label{eq:heisenberg}
\end{equation}  
where ${\cal L}_{D}$ is a generator of a completely positive Liouville super operator.

\begin{figure}[htbp]
\vspace{1cm}
\hspace{2cm}
\center{\includegraphics[height=7cm]{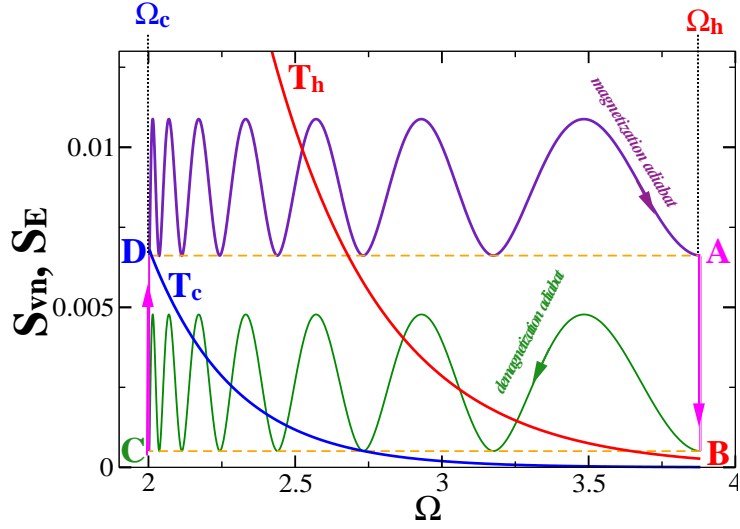}}
\caption{Refrigerator cycle in the frequency entropy plane. 
The von Neumann entropy ${\cal S}_{vn}=-tr\{ {\Op \rho}  \log {\Op \rho} \}$ ({\bf ABCD} rectangle)
as well as the energy entropy ${\cal S}_E=-\sum{ p_i \log p_i}$ are shown
($p_i$ is the population of energy level $i$). 
The hot and cold isotherms are indicated. On the {\em adiabats} the energy level spacings change from $\Omega_h$
to $\Omega_c$
The demagnetization {\em adiabat} and the magnetization {\em adiabats} revolve exactly seven periods.
On the {\em isochores} the energy level spacing remains constant and the entropy changes due to change
in population.  The cycle parameters are:  $J=2$, $T_c=0.24 $, $T_h=1.18$, $\omega_c=0.1$, $\omega_h=3.325$, 
$\tau_c=10.54$, $\tau_h=9.741$, $\tau_{hc}=\tau_{ch}=12.81$.}
\label{fig:1}
\end{figure}

The cycle studied is composed of two isomagnetic segments where the working medium is 
in contact with the cold/hot baths and the external control field $\omega$ is constant, termed 
{\em isochores}. In addition, there are two segments  termed {\em adiabats} where the external field $\omega(t)$ varies and with it the energy level structure of 
the working medium. This cycle is a quantum analogue of the 
Otto cycle \cite{k176}. Each segment is characterized by a quantum propagator
${\cal U}_s$. The propagator
maps the initial state of the working medium to the final state 
on the relevant segment.
The four strokes of the cycle in analogy with the Otto cycle  (see Fig. \ref{fig:1}  ) are:

\begin{itemize}
\item{ {\em hot  isomagnetic (Isochore)} $A \rightarrow B$: the field is maintained
constant $\omega=\omega_h$ the working medium
is in contact with the hot bath of temperature $T_h$.
${\cal L}_D$ leads to equilibrium  with heat
conductance $\Gamma_h$, for a period of $\tau_h$.  The segment dynamics is described by the
propagator ${\cal U}_h $. }
\item{ {\em demagnetization (expansion) adiabat} $B \rightarrow C$: The field changes
from $\omega_h$ to $\omega_c$ in a time period  of $\tau_{hc}$.
${\cal L}_D= {\cal L}_N$ represents external noise in the controls.
The propagator becomes  ${\cal U}_{hc}$ which is the main subject of study.}
\item{ {\em cold isomagnetic (Isochore)} $C \rightarrow D$: the field
is maintained  constant $\omega=\omega_c$ the working medium
 is in contact with the cold bath of temperature $T_c$. ${\cal L}_D$ leads to equilibrium  with heat
conductance $\Gamma_c$, for a period of $\tau_c$.  The segment dynamics is described by the
propagator ${\cal U}_c $.      }
\item{ {\em Magnetization (compression) adiabat} $D \rightarrow A$: The field  changes
from $\omega_c$ to $\omega_h$ in a time period of $\tau_{ch}$, 
${\cal L}_D= {\cal L}_N$ represents external noise in the controls.
The propagator becomes  ${\cal U}_{ch}$.}
\end{itemize}

The product of the four propagators, ${\cal U}_{s}$  is the cycle propagator: 
\begin{equation}
{\cal U}_{cyc}~~=~~{\cal U}_{ch} {\cal U}_c {\cal U}_{hc} {\cal U}_h ~. 
\label{eq:globop}
\end{equation} 
Eventually, independent of initial condition, after a few cycles, the working medium 
will reach a limit cycle characterized as  
an invariant eigenvector of ${\cal U}_{cyc} $ with eigenvalue $\bf 1$(one) \cite{k201}. 
The characteristics of the refrigerator are therefore extracted from the limit cycle.

\section{Quantum thermodynamical observables and their dynamics} 

To facilitate the study of the dynamics of the cooling cycle we need a representation of the state $\Op \rho$
and the thermodynamical observables.
The orthogonal set of  time independent operators  $ {\bf \hat B_i} $, are closed to the dynamics.
As a result they can supply a complete vector space to expand the propagators ${\cal U}$ and $\Op \rho$.
A thermodynamically oriented time dependent vector space  which directly addresses the issue of adiabaticity 
is superior. This set includes the energy $ {\bf \hat H} $ and two additional orthogonal operators:
\begin{equation}
{\bf \hat H} \rm~ ~=~\omega (t) {\bf \hat B_1 } \rm~+~J {\bf \hat B_2 }
 \rm
~~,~~
 {\bf \hat L}  \rm~~=~-J {\bf \hat B_1 } \rm~+~\omega(t) {\bf \hat B_2}
 \rm
~~,~~
 {\bf \hat C} \rm ~ ~=~  \Omega(t)    {\bf \hat B_3 } \rm
\label{defnew} 
\end{equation} 
To uniquely define the state of the system $\Op \rho$ the original set is supplemented with two  operators:
${\Op V} = \Omega {\Op B_4}= \frac{1}{2}\Omega ({\Op I}^1 \otimes {\Op \sigma}_z^2 -{\Op I}^2 \otimes {\Op \sigma}_z^1)$ and 
${\Op D} = \Omega {\Op B_5}=\Omega {\Op \sigma}_z^1 \otimes {\Op \sigma}_z^2$. With this operator base the state $\Op \rho$ can be expanded as:
\begin{equation}
\Op \rho =\frac{1}{4} {\Op I} + \frac{1}{ \Omega} \left( \langle \Op H \rangle \Op H +\langle \Op L \rangle \Op L 
+\langle \Op C \rangle \Op C  + \langle \Op V \rangle \Op V +\langle \Op D \rangle \Op D  \right) 
\label{eq:rho}
\end{equation}
$\Op V$ and $\Op D$ commute with $\Op H$. The equilibrium value of $\langle \Op V \rangle$ is  zero,
and once it reaches equilibrium it does not change during the cycle dynamics. 
As a result the state $\Op \rho $ can be described in the energy representation by four expectation values:
\begin{eqnarray}
\Op \rho_e = \frac{1}{4} \left(
\begin{array}{cccc}
1+ \frac{4}{\Omega}(D -E)&0&0&\frac{4}{\Omega}(L+iC)\\
0&1- \frac{4}{\Omega}D &0&0\\
0&0&1- \frac{4}{\Omega}D &0\\
\frac{4}{\Omega}(L-iC)&0&0&1+ \frac{4}{\Omega}(D+E)
\end{array}
\right)
\label{eq:rhoe}
\end{eqnarray}
where $E = \langle \Op H \rangle $,  $L = \langle \Op L \rangle $, $C = \langle \Op C \rangle $ and $D = \langle \Op D \rangle $.
From Eq. (\ref{eq:rhoe}) it is clear that when $L=C=0$, $\Op \rho_e $ is diagonal in the energy representation, then
$[\Op \rho(t), \Op H(t) ] =0$.
It therefore can be concluded that in complete adiabatic following $L$ and $C$ are maintained at zero value.

\subsection{The dynamics on the {\em adiabats}.}

In general the dynamics on the demagnetization {\em adiabat} is generated by ${\cal L} ={\cal L}_H + {\cal L}_N$ 
where ${\cal L}_H= \frac{i }{\hbar}[\Op H, \cdot ]~$ and $\Op H(t)$ is the time dependent Hamiltonian,   Eq. (\ref{eq:hamil}).  
The external noise generator ${\cal L}_N$ is defined later.  
The equations of motions for the dynamical observables
$ {\bf \hat B_1} ,~ { \bf \hat B_2} ,
~ { \bf \hat B_3}  $~~
become:
\begin{eqnarray}
\frac{d}{dt} \left( \begin{array}{c}
 {\bf \hat B_1 }  \\
 {\bf \hat B_2 }  \\
{\bf \hat B_3 } \\
\end{array} \right)(t)=
\left(
\begin{array}{ccc}
0  & 0  & J  \\
0  & 0  & -\omega  \\
-J  &  \omega & 0  \\
\end{array}
\right) \left(\begin{array}{c}
{\bf \hat B_1 }  \\
{\bf \hat B_2}\ \\
{\bf \hat B_3}  \\
\end{array} \right)
\label{oldad} 
\end{eqnarray}
Our purpose is to evaluate the deviation from perfect adiabatic following. 
The equation of motion for the time dependent set $  {\Op H } $, 
$ {\Op L }$  and $ {\Op C } $ 
leading to the propagators ${\cal U}_{hc}$ and ${\cal U}_{ch}$ are appropriate for this task.
The integration to obtain ${\cal U}_{hc}$ and ${\cal U}_{ch}$ will be carried out  with respect to a new time variable $d \Theta = \Omega dt $:
\begin{eqnarray}
\frac{d}{\Omega dt} \left( \begin{array}{c}
 {\bf \hat H }  \\
 {\bf \hat L }  \\
 {\bf \hat C }  \\
\end{array} \right)(t)=
\left(
\begin{array}{ccc}
\frac {\dot \Omega} {\Omega^2}   & - \frac {J \dot \omega}{\Omega^3}   & 0  \\
 \frac {J \dot \omega}{\Omega^3} & \frac {\dot \Omega } {\Omega^2}   &-1   \\
0  & 1 & \frac {\dot \Omega} {\Omega^2}         \\
\end{array}
\right) \left(\begin{array}{c}
{\bf \hat H } \\
{\bf \hat L}\\
{\bf \hat C}  \\
\end{array} \right)
\label{eq:newanad1} 
\end{eqnarray} 
The ability of the working medium to follow the energy spectrum is defined by the adiabatic measure:
\begin{equation} 
\mu =  \frac {J \dot \omega}{\Omega^3} 
\label{eq:m}
\end{equation}
We find that $\mu$ is a major parameter that characterizes the dynamics on the {\em adiabats}.
When $\mu=0$ the propagator factorizes, the dynamics of $\Op H$ is independent of $\Op L$ and $\Op C$. 
A large $\mu$ will cause large non-adiabatic changes coupling $\Op H$ with $\Op L$ and $\Op C$.
We will show that constant $\mu$ minimizes the accumulated nonadiabatic transitions.

Another nice feature of constant  $\mu$ is that  Eq. (\ref{eq:newanad1}) can be integrated
leading to a closed form solution for the demagnetization and magnetization propagators ${\cal U}_{hc}$ 
and ${\cal U}_{ch}$.
The consequence  of stationary $\mu$ is a particular scheduling function of the external field $\omega(t)$ with time:
\begin{equation}
\omega(t) =\frac{J {f}}{\sqrt{1-f^2}}~~~,~~~\Omega(t)=\frac{J}{\sqrt{1-f^2}}
\end{equation}
where $f(t)$ is a linear function of time:
$f_{hc}(t)=\frac{\omega(t)}{\Omega(t)}=\frac {t}{\tau_{hc}} \left(
  \frac {\omega_c}{\Omega_c} ~-~ \frac {\omega_h}{\Omega_h}   \right) ~+~
\frac {\omega_h}{\Omega_h} $.
Swapping $h$ for $c$ in $f(t)$ leads to the equivalent expression
for the magnetization {\em adiabat}.

The adiabatic parameter and the time allocated to the {\em adiabat}
obey the reciprocal relation: $ \mu_{hc}~=~ \frac{K_{hc}}{\tau_{hc}} $
where $ K_{hc}= \frac{1}{J} \left(\frac {\omega_c}{\Omega_c} ~-~ \frac {\omega_h}{\Omega_h} \right) $.
Swapping $c$ with $h$ leads to  $ \mu_{ch }$ and then $K_{ch}=-K_{hc}$.

The solution is facilliated the time variable $\Theta$, $d \Theta = \Omega dt$.
The final values of $\Theta_{hc} $  becomes: $ \Theta_{hc }   ~ = ~ \tau_{hc} \frac{1}{K_{hc}} \Phi_{hc}$\\
where:
$
\Phi_{hc}~  = ~
\left( \arcsin(
\frac{\omega_c}{\Omega_c}) -\arcsin(\frac {\omega_h}{ \Omega_h}) \right)
$
and $ 0 \ge \Phi \ge -\frac{\pi}{2}$. 

Eq. (\ref{eq:newanad1})  is solved by noticing that the diagonal part is a unit matrix
multiplied by a time dependent scalar. Therefore we seek a solution of the type 
${\cal U}_{hc}= {\cal U}_1 {\cal U}_2$ where $[ {\cal U}_1, {\cal U}_2] =0$.
The integral of the diagonal part of  Eq. (\ref{eq:newanad1})  becomes: 
\begin{equation}
{\cal U}_1 ~~=~~e^{(\int_0^{\tau_{hc}} \frac {\dot{ \Omega}}{\Omega}dt)} \Op 1~~=~~ 
\frac {\Omega_c}{\Omega_h} \Op 1~~,
\label{eq:soldiag} 
\end{equation} 
which can be interpreted as the scaling of the energy levels with the variation in $\Omega$.

To integrate ${\cal U}_2$ the non diagonal parts of Eq. (\ref{eq:newanad1}),  are diagonalized, 
leading to the eigenvalues $0,-i \sqrt{q}, i \sqrt{q}$, where $q= \sqrt{1 +\mu^2}$,
and the propagator:
\begin{eqnarray}
{\cal U}_2~~=~~
\left(
\begin{array}{ccc}\frac {1 + \mu^2 c}{q^2}   & -\frac{\mu s}{q}   &  \frac {\mu(1-c)}
{q^2}   \\
 \frac {\mu s} { q}       & c    &- \frac {s}{q}   \\
 \frac {\mu (1-c)}{ q^2}  & \frac {s}{q} & \frac {\mu^2+c} {q^2}  \\
\end{array}
\right) ~~,
\label{eq:calprop}
\end{eqnarray} 
where $s=sin(q \Theta)$ and $c=cos(q \Theta)$.
The propagator ${\cal U}_2$ induces periodic mixing of $  {\Op H } $ with $  {\Op L } $  and $  {\Op C } $.  
As a result  a diagonal $\Op \rho_e$ Cf. Eq. (\ref{eq:rhoe}) will develop non diagonal terms.
To characterize the deviation from perfect factorization of $  {\Op H } $ from $  {\Op L } $  and $  {\Op C } $,
we define an adiabaticity measure  $\delta$ as: 
\begin{equation}
\delta=1-{\cal U}_1^{-1}{\cal U}_{hc}(1,1)~~. 
\label{eq:delta}
\end{equation}
where ${\cal U}_1^{-1}=\frac{\Omega_h}{\Omega_c} \Op 1$ is introduced to correct for the energy scaling.
In the present context of noiseless dynamics and constant $\mu$, $\delta = 1 - {\cal U}_2(1,1)$. 
When $\delta=0$ there is complete factorization.
As will be described in Sec. \ref{sec:thermodyn} $\delta \ne 0$ determines the minimum temperature.
 
The adiabatic limit is described by $\mu \rightarrow 0$. Then  Eq.  (\ref{eq:calprop}) converges to the identity operator. 
These are the perfect adiabatic following conditions where $\delta=0$.
In general Eq. (\ref{eq:calprop}) describes a periodic motion of $\Op H$ $\Op L$ and $\Op C$.
Each period  is defined by
\begin{equation}
q \Theta  =  ~2 \pi l~~l=0, 1,2 ...
\label{eq:mfric} 
\end{equation}  
where  $l$ is the winding number. At the end of each period  ${\cal U}_2$ restores to the identity matrix. 
These are the periodic frictionless conditions where $\delta=0$. For intermediate times 
$\langle {\Op H} \rangle$ is always larger than the frictionless value $\delta > 0$. 
The amplitude of this periodic dynamics decreases when $\mu$ becomes smaller, Cf. 
${\cal U}_2(1,1)$ in Eq. (\ref{eq:calprop}). 
Constant $\mu$ is the minimum of $\delta$ Eq. (\ref{eq:delta}) (Cf. appendix \ref{app:A}

The frictionless conditions define a quantization condition for 
the adiabatic parameter $\mu$:
\begin{equation}
\mu ~~=~~ \left(~\left(\frac{2 \pi l}{\Phi_{hc}} \right)^2  -1 \right)^{-\frac{1}{2}}~~.
\label{eq:mquant}
\end{equation}
Examining Eq. (\ref{eq:mquant}) we find that there is no solution for $l=0$.
The first frictionless solution $l ~\ge ~\sqrt{\frac{\Phi_{hc}}{2 \pi}}$ leads to a minimum demagnetization time:
\begin{equation}
\tau_{hc}(min) = K_{hc}
\sqrt{ \left(\frac{2 \pi}{\Phi_{hc}} \right)^2  -1}~~.
\label{eq:taumin}
\end{equation}
From Eq. (\ref{eq:taumin}) we can interpret that the minimal frictionless demagnetization time scales as 
$\tau_{hc}(min) \propto \frac{1}{J}$, since it has a weak dependence on $\omega_c$ and $\omega_h$ .
The special closed form solution can be employed in a piecewise fashion to analyze other scheduling functions $\omega(t)$. In general we expect similar quantization of the solutions.
The main observation of this section is that we can find families of periodic  frictionless solutions where the energy 
restores to its adiabatic value every period. For $\mu \rightarrow 0 $ these solutions coalesce with the adiabatic following solutions.
Table I summarizes some of the notations used.

\subsection{The influence of noise }
\label{existmint}

Any realistic refrigerator is subject to noise on the external controls. The main point of this paper is that
even an infinitesimal amount of noise will eliminate the frictionless solutions.
The sensitivity to noise results from the  requirement of precise control of the scheduling of the external field $\omega(t)$. 
To observe this effect requires a model of the noise induced by  the external controls.

First we consider a piecewise process controlling the scheduling of $\omega$ in time. At every time interval,
$\omega$ is updated to its new value. For such a procedure random errors are expected in the duration of these time intervals   described by the Liouville operator ${\cal L}_N$. We model these errors as a Gaussian delta correlated noise.
This process is  mathematically equivalent to a dephasing process on the demagnetization {\em adiabat} \cite{k215}.  
This  stochastic dynamics can be modeled by a Gaussian semigroup with the generator \cite{gorini76,breuer}:
\begin{equation}
{\cal L}_{N_p} (\Op A)~~=~~ -\frac{\gamma_p}{\hbar^2} [ {\Op H}, [ {\Op H}, {\Op A} ]]~~,
\label{eq:h1n}
\end{equation}
which is termed phase noise, Eq. (\ref{eq:h1n}). An equivalent dynamics to Eq. (\ref{eq:h1n})  is also obtained 
in the limit of weak quantum measurement of the instantaneous energy \cite{lajos06}. 
For this noise model the modified equations of motion on the {\em adiabats} become:
\begin{eqnarray}
\frac{d}{\Omega dt} \left( \begin{array}{c}
 {\bf \hat H }\\
 {\bf \hat L } \\
  {\bf \hat C }  \\
\end{array} \right)(t)=
\left(
\begin{array}{ccc}
\frac {\dot \Omega} {\Omega^2}   & - \frac {J \dot \omega}{\Omega^3}   & 0  \\
 \frac {J \dot \omega}{\Omega^3} & \frac {\dot \Omega } {\Omega^2} -{\gamma_p}{\Omega}  &-1   \\
0  & 1 & \frac {\dot \Omega} {\Omega^2} -{\gamma_p}{\Omega}        \\
\end{array}
\right) \left(\begin{array}{c}
{\bf \hat H } \\
{\bf \hat L} \\
{\bf \hat C}  \\
\end{array} \right)~~.
\label{eq:newanad} 
\end{eqnarray} 
The term $ {\gamma_p}$  describing the phase noise is assumed to be small. We therefore seek a perturbative solution: ${\cal U}_a~=~{\cal U}_1 {\cal U}_2 {\cal U}_3 $ where  the equations of motion ${\cal U}_3$ can be obtained from the interaction representation:
\begin{eqnarray}
\label{eq:u3} 
\frac{d }{ \Omega dt} {\cal U}_3 (t)~~=~~ {\cal U}_2(-t) \left(
\begin{array}{ccc}
0 & 0& 0\\
0 &- {\gamma_p}{\Omega} & 0 \\
0 & 0 & - {\gamma_p}{\Omega}
\end{array} 
\right) {\cal U}_2 (t) ~{\cal U}_3(t) ~=~{\cal W}(t) {\cal U}_3(t) 
\end{eqnarray}
where:
\begin{eqnarray}
{\cal W}(t)=
\gamma_p \Omega(t) \left(
\begin{array}{ccc}
\frac{\mu^2}{q^4}(s^2 \mu^2+2(1-c)) & \frac{\mu s}{q^3}(\mu^2c+1)& -\frac{\mu}{q^4}(1-c)(\mu^2c+1)\\
\frac{\mu s}{q^3}(\mu^2c+1) &\frac{\mu^2 c^2+1}{q^2}& \frac{\mu^2s}{q^3}(1-c) \\
-\frac{\mu}{q^4}(1-c)(\mu^2c+1) & \frac{\mu^2s}{q^3}(1-c) &1-\frac{\mu^2}{q^4}(1-c)^2
\end{array} \right) 
\label{eq:wmatrix}
\end{eqnarray}
${\cal U}_3$ describes the dynamics with respect to the reference provided by the unitary trajectory ${\cal U}_2$.
We seek an approximate solution for ${\cal U}_3$ in the limit
when $\mu \rightarrow 0$, then  ${\cal U}_2 = {\cal I}$ since this is the frictionless limit.
Expanding Eq. (\ref{eq:wmatrix}) to first order in $\mu$ leads to:
\begin{eqnarray}
{\cal W}(t) \approx
\gamma_p \Omega(t) \left(
\begin{array}{ccc}
0& \mu s& \mu(1-c)\\
\mu s&1& 0\\
\mu (1-c)& 0&1
\end{array} \right) 
\end{eqnarray}
${\cal U}_3(\tau_{hc})$ is solved in two steps.  First evaluating the propagator for one period of $\Theta$:
for which $\Omega(t)$ is almost constant, and then the global propagator
becomes the product of the one period propagators for $l$ periods:
${\cal U}_3(\tau_{hc})\approx {\cal U}_3(\Theta=2 \pi)^l $. 
The Magnus expansion \cite{magnus} to second order is employed to obtain the one period propagator ${\cal U}_3(2 \pi )$:
\begin{equation}
{\cal U}_3(\Theta=2 \pi) ~~\approx e^{ {\cal M}_1 +{\cal M}_2 + ...}
\end{equation}
where: 
${\cal M}_1=\int_0^{2 \pi} d \Theta W(\Theta) $ and
${\cal M}_2=\frac{1}{2} \int_0^{2 \pi} \int_0^{\Theta} d \Theta d\Theta' [ {\cal W}(\Theta), {\cal W} (\Theta')]  + ...)$
The first order Magnus term leads to:
\begin{eqnarray}
{\cal U}_3 (\Theta=2 \pi)_{M_1} \approx
 \left(
\begin{array}{ccc}
1& 0& \mu(1-e^{- 2 \pi \gamma_p \Omega})\\
0&e^{- 2 \pi \gamma_p \Omega} & 0\\
\mu(1-e^{- 2 \pi \gamma_p \Omega})& 0&e^{- 2 \pi \gamma_p \Omega} 
\end{array} \right) 
\end{eqnarray}
which to first order in $\mu$, $\delta$ stays zero. ${\cal U}_{3}$ does not couple
$\langle \Op H \rangle$ with $\langle \Op L \rangle$ and $\langle \Op C \rangle$.

The second order Magnus approximation leads to:
\begin{eqnarray}
{\cal U}_3 (\Theta=2 \pi)_{M_2} \approx
 \left(
\begin{array}{ccc}
C& -S& 0\\
S&C& 0\\
0& 0&1
\end{array} \right) 
\label{eq:sivuv}
\end{eqnarray}
where $S= \sin \alpha$ and $C=\cos\alpha$.  
$\alpha = \gamma_p \Omega \pi \mu \sqrt{9 \mu^2+4}$ and as $\mu^2 \ll \frac{4}{9}  $,
$\alpha = 2 \gamma_p \Omega \pi \mu \approx  \Phi_{hc} \gamma_p \Omega \frac{1}{l}$, Cf. Eq. (\ref{eq:mquant}).
The condition $\mu^2 \ll \frac{4}{9}  $ can be transformed to $ l \gg \frac{9 \Phi}{8 \pi} $.  
An {\em adiabat} with a small number of revolutions $l \approx 10$ already fulfills this condition.
We now combine the second order propagator  ${\cal U}_3(\tau_{hc})$, for $l$ revolutions. It has also the structure 
of a rotation matrix identical to
Eq. (\ref{eq:sivuv}), with  a new  rotation angle $\alpha=\alpha_l$, where:
\begin{equation}
\alpha_l =   2 \pi \gamma_p \mu \int_0^{2 \pi l} \Omega(\Theta) d \Theta = \pi \gamma_p J \ln\left[\frac{(\Omega_h+\omega_h)(\Omega_c-\omega_c)}{(\Omega_h-\omega_h)(\Omega_c +\omega_c)}\right]
\label{eq:alphal}
\end{equation}
The asymptotic value of $\alpha_l$ is finite when $\mu \rightarrow 0$.

For the quantization conditions when ${\cal U}_2=\Op 1$ the deviation of ${\cal U}_3$ 
from the identity operator defines $\delta$. 
Asymptotically as $\mu \rightarrow 0~$ and $\omega_c \ll J$,
\begin{equation} 
\delta_{min} = 1-\cos (\alpha_l)  \approx  \pi^2 \gamma_p^2 J^2  \ln[\omega_h/J] 
\label{eq:phasnoise}
\end{equation}

Another source of external noise is induced by fluctuations in the
frequency $\omega(t)$. 
Such a term represent Markovian random fluctuations in the external magnetic filed.
If the fluctuations are fast compared to $2 \pi / \Omega$, such noise can be described by the Lindblad term: 
${\cal L}_{\omega} \Op X = - \gamma_a \omega^2 [ \Op B_1 , [ \Op B_1, \Op X]]$.
\begin{eqnarray}
\label{eq:u3} 
\frac{d }{ \Omega dt} {\cal U}_3 (t)~~=~~ -\gamma_a \frac{\omega^2}{\Omega} {\cal U}_2(-t) \left(
\begin{array}{ccc}
 \frac{J^2}{\Omega^2}& \frac{J \omega}{\Omega^2}& 0\\
\frac{J \omega}{\Omega^2} &\frac{\omega^2}{\Omega^2}& 0 \\
0 & 0 & 1
\end{array} 
\right) {\cal U}_2 (t) ~{\cal U}_3(t) ~=~{\cal W}(t) {\cal U}_3(t) 
\end{eqnarray}
We seek an approximate solution for the quasistatic limit
when $\mu \rightarrow 0$. Expanding ${\cal W}$ in Eq. (\ref{eq:u3}) to zero order in $\mu$ leads to:
\begin{eqnarray}
{\cal W}(\Theta) \approx
 -\gamma_a \frac{\omega^2}{\Omega^3 } \left(
\begin{array}{ccc}
J^2& J \omega c& -J \omega s\\
J \omega c&\omega^2+s^2J^2& sc J^2\\
-J \omega s& s c J^2&J^2c^2+\omega^2
\end{array} \right) 
\end{eqnarray}
We calculating the propagator for an integer number of periods the lowest order Magnus expansion becomes:
$ {\cal U}_3 (\Theta=2 \pi l) ~~=~~ \exp \left( {\int_0^{2 pi l} d \Theta {\cal W}(\Theta)} \right) $,
then  the ${\cal U}_3(1,1)$ element decouples from the remaining part of the propagator and becomes:
\begin{equation}
{\cal U}_3(1,1) = \exp \left[ -\gamma J^2 \int_0^{2 \pi l} d \Theta \left (\frac{\omega^2(\Theta)}{\Omega^3(\Theta)} \right) \right]
\label{eq:u311}
\end{equation}
Eq. (\ref{eq:u311}) can be integrated and since  ${\cal U}_2=\Op 1$ for an integer number of revolutions then:
\begin{equation}
\delta = 1-{\cal U}_3(1,1)~~ \approx ~~1-e^{-\gamma_a   \frac{J^2 \omega_h^2 }{3 \Omega_c^2} \tau_{hc}}~~.
\label{eq:ampnoise}
\end{equation}
The smallest $\delta$ is achieved for a one period cycle, Eq. (\ref{eq:taumin}) then:
$ \delta_{min} \approx \gamma_a J  \frac{4 \omega_h^2}{\Omega_h^2} ~~$.
The phase noise and the amplitude noise have a reciprocal relation with respect to $l$.
Phase noise is maximized for small $l$ and amplitude noise for large $l$.
Another possible source of noise is caused by fluctuation in the interaction energy $\Op H_{int}$.
Analysis shows that such noise will lead to a similar expression to Eq. (\ref{eq:ampnoise}) where $J^2$
is replaced by $\omega_c^2$.

\subsection{The dynamics on the isomagnetic segments ({\em isochores})}

On the {\em isochores} the equation of motion lead to equilibration with the hot and cold baths respectively.
During the process the Hamiltonian is constant which leads to a factorization of the equations
of motion:
\begin{eqnarray}
\frac{d}{dt} \left( 
\begin{array}{c}
\Op H\\
\Op L\\
\Op C\\
\Op D \\
\Op I\\
\end{array} \right)
~~=~~ \left(
\begin{array}{ccccc}
-\Gamma&0&0&0&\Gamma E_{eq}\\
0&-(\Gamma+\gamma \Omega^2)&-\Omega&0&0\\
0&\Omega&-(\Gamma+\gamma \Omega^2)&0&0\\
\frac{2}{\Omega} \Gamma E_{eq}&0&0&-2\Gamma&0\\
0&0&0&0&0\\
\end{array} \right)
\left( 
\begin{array}{c}
\Op H\\
\Op L\\
\Op C\\
\Op D\\
\Op I\\
\end{array} \right)
\label{eq:dyniso}
\end{eqnarray}
where $\Gamma = \kappa^+ + \kappa^-$, $\frac{\kappa^+}{\kappa^- }=e^{-\frac{\hbar \Omega}{k_B T}}$
and $E_{eq}= \hbar \Omega(\kappa^+ - \kappa^-)/\Gamma= \hbar \Omega ( e^{-\frac{2 \hbar \Omega}{k_B T}}-1)/Z $,
where $Z= 1+2e^{-\frac{\hbar \Omega}{k_B T}} +e^{-\frac{2 \hbar \Omega}{k_B T}}$.
In Eq. (\ref{eq:dyniso}) $\Op H$ decouples  from $\Op L$ and $\Op C$.

Eq. (\ref{eq:dyniso}) can be integrated leading to:
\begin{eqnarray}
\begin{array}{l}
\Op H (t) ~~=~~  e^{-\Gamma t}(\Op H (0) - E_{eq} \Op I ) \\
\Op L (t) ~~=~~ e^{-\Gamma t}(\Op L (0) \cos \Omega t - \Op C(0) \sin \Omega t )\\
\Op C (t) ~~=~~ e^{-\Gamma t}(\Op C(0) \cos \Omega t +  \Op L(0) \sin \Omega t ) \\
\Op D(t) ~~=~~\Op D(0) e^{-2 \Gamma t} + \frac{1}{\Omega} \left(\Op H(0)E_{eq}(e^{-\Gamma t} - e^{-2 \Gamma t} ) -E_{eq}^2 (e^{-\Gamma t} -1) \Op I \right)
\end{array}
\label{eq:solved}
\end{eqnarray}
From Eq. (\ref{eq:solved}) the propagators ${\cal U}_{c}$ and ${\cal U}_h$ can be constructed.

\begin{table}
\caption{Notation and definitions} 
\begin{center}
\begin{tabular}{||c|c|c||}
\tableline
Name &  Notation & Comments \\
\tableline
Compression ratio&${\cal C}$&${\cal C}= \frac{\Omega_h}{\Omega_c}$\\
\tableline
Reversibility&${\cal R}$&${\cal R}=\frac{T_c \Omega_h}{T_h \Omega_c}$\\
\tableline
Adiabatic measure&$\mu$&$\mu =  \frac {J \dot \omega}{\Omega^3}$ \\
\tableline
Reciprocal relation&$K=\tau \mu$&$K_{hc}=\frac{1}{J}\left(\frac{\omega_c}{\Omega_c}-\frac{\omega_h}{\Omega_h}\right)$\\
\tableline
Compression angle&$\Phi$&$\Phi_{hc}=
\left( \arcsin(\frac{\omega_c}{\Omega_c})-\arcsin(\frac{\omega_h}{\Omega_h})\right)$\\
\tableline
Rotation angle&$\Theta$&$\Theta = \frac{\Phi}{\mu}$\\
\tableline
Heat conductivity&$\Gamma=\kappa^+ + \kappa_-$&$\frac{\kappa^+}{\kappa_-}=e^{-\frac{\hbar \Omega}{k_B T}}$\\
\tableline
Adiabaticity &$\delta$&$\delta= 1 - {\cal U}_1^{-1} {\cal U}_{hc}$\\
\tableline
Phase noise &$\gamma_p$&$-\frac{\gamma_p}{\hbar^2} [ {\Op H}, [ {\Op H}, {\Op A} ]]$\\
\tableline
Amplitude noise &$\gamma_a$&$-{\gamma_a}{\omega^2} [ {\Op B}_1, [ {\Op B}_1, {\Op A} ]]$\\
\tableline
\end{tabular}
\end{center}
\label{tab:notations}
\end{table}

\section{Thermodynamical relations}
\label{sec:thermodyn}

The maximal efficiency $ \eta^{max} $,  of a heat engine is 
limited by the second law to the  Carnot efficiency.
For the quantum Otto type cycle the efficiency is limited by the ratio of the energy level difference in the hot and cold sides \cite{k190,mahler07}. 
As a result we obtain the series of inequalities:   
\begin{equation}
\eta^{max}~=~1~-~ \frac  {\Omega_c} {\Omega_h}~~<~1~-~   \frac {\omega_c} {\omega_h}~~ ~<~~ 1~~-~~\frac {T_c}{T_h} 
\label{eq:heatineq}   
\end{equation} 
In the operation as a refrigerator the inequality in Eq. (\ref{eq:heatineq}) is reversed. 
This imposes a restriction on the minimum cold bath temperature $T_c$:
\begin{equation} 
 {T_c}~~\ge~~\frac  {\Omega_c} {\Omega_h} T_h~ ~, 
\label{eq:basicineq}   
\end{equation} 
$\Omega_c$ is limited by $J$ and for the limit $\omega_h \gg J$ we obtain:
\begin{equation} 
 {T_c}~~\ge~~\frac  {J} {\omega_h} T_h~ ~,
\label{eq:basicineq}   
\end{equation}  
On the cold side the necessary condition for refrigeration is that the internal 
energy of the working medium at the end of the demagnetization
is smaller than the equilibrium energy with the cold bath (Cf. Fig. \ref{fig:1} and \ref{fig:typical}).
\begin{equation}
\langle \Op H \rangle_C ~\le~\langle \Op H \rangle_{eq} (T_c) = - \hbar \Omega_c \left(1 -  2 e^{- \frac{\hbar \Omega_c}{k_b T_c}} \right)~,
\label{eq:fff}
\end{equation}
where $\langle \Op H \rangle_{eq}(T_c)$ is approximated by the low temperature limit $ \hbar \Omega_c \gg k_B T_c$.
On the hot {\em isochore} the lowest energy  point $B$, that  can be  obtained, is in equilibrium with $T_h$:\\
$\langle \Op H \rangle_B \ge \langle \Op H \rangle_{eq} (T_h) = - \hbar \Omega_h \left(1 -  2 e^{- \frac{\hbar \Omega_h}{k_b T_h}} \right)~$. Under these conditions $ L=C=0$.
The change in $\langle \Op H \rangle$ in the demagnetization {\em adiabat} leads to:
\begin{equation}
\langle \Op H \rangle_C \approx  \frac{ \Omega_c}{\Omega_h} \left(1 -  \delta \right) \langle \Op H \rangle_B~~.
\label{eq:hcmin}
\end{equation}
where $\delta$ the deviation from frictionless solutions is  defined in Eq. (\ref{eq:delta}).
Then the maximum heat that can be extracted per cycle becomes:
\begin{equation}
{\cal Q}_c (max) ~~=~~ \langle \Op H \rangle_{eq}(T_c) -\langle \Op H \rangle_C \approx 2 \hbar \Omega_c \left(e^{- \frac{\hbar \Omega_c}{k_b T_c}} -e^{- \frac{\hbar \Omega_h }{k_bT_h}} - \frac{1}{2}\delta~\right) 
\label{eq:maxq}
\end{equation}
The condition for refrigeration is ${\cal Q}_c (max) \ge 0$. When $\delta \ll e^{- \frac{\hbar \Omega_h }{k_bT_h}}$
the minimum temperature becomes the Carnot limit Eq. (\ref{eq:basicineq}). For sufficiently large
$\omega_h $, positive ${\cal Q}_c (max) \ge 0$ leads to
$ \delta \le 2 e^{-\frac{\hbar \Omega_c}{k_b T_c}} $ imposing a
stronger restriction on the minimal temperature:
\begin{equation}
T_c \ge \frac{\hbar J}{-k_B \log (\delta/2) }
\label{eq:mintemp}
\end{equation}
Due to the logarithmic dependence on the noise $\delta$ the minimum temperature scales linearly with the energy gap $\hbar J$. Eq. (\ref{eq:mintemp}) relates the minimum temperature to $\delta$ the adiabticity parameter.

\section{Power optimization}
\label{sec:power}

The cooling power ${\cal P}_c$ is the amount of heat extracted ${\cal Q}_c$ divided by the cycle time $\tau$.
For  the frictionless solutions on the {\em adiabats} the heat extracted  is obtained by considering the balance of heat and work required to close the cycle \cite{k116,k221}:
\begin{equation}
{\cal Q}_c = \hbar \Omega_c (\frac{E_{eq}^{h}}{\Omega_h} -\frac{E_{eq}^{c}}{\Omega_c})\frac{(e^{x_{c}}-1)(e^{x_{h}}-1)}
{1-e^{x_{c}+x_{h}}}
 \approx  2\hbar \Omega_c \left(e^{-\frac{\hbar \Omega_c}{k_b T_c}}-e^{-\frac{\hbar \Omega_c}{k_b T_c}}\right)F(x_{c},x_{h})~~,
\end{equation}
where: $x_c= \Gamma_c \tau_c$ and $x_h =\Gamma_h \tau_h$.
Optimizing the cooling power becomes equivalent to optimizing $\frac{F(x_c,x_h)}{\tau_{cyc}}$
where $\tau_{cyc}=\tau_h +\tau_{hc}+\tau_c +\tau_{ch}$ is the total cycle time. 
For frictionless solutions the minimum time on the {\em adiabats }$\tau_{hc}$ and $\tau_{ch}$ is described in Eq. (\ref{eq:taumin}).
The optimal partitioning of the time allocation between the hot and cold {\em isochores}
is obtained when:
\begin{equation}
\Gamma_{h}(\cosh(\Gamma_{c}\tau_{c}) -1 )=\Gamma_{c}(\cosh(\Gamma_{h}\tau_{h})-1)~~.
\label{eq: d-tau optimum work}
\end{equation}
When $\Gamma_{h}=\Gamma_{c}$
the optimal time allocations on the {\em isochores} becomes  $\tau_{h}=\tau_{c}$. 

The total time allocation $\tau=\tau_{iso}+\tau_{adi}$ is partitioned to the time on the
{\em adiabat}s $\tau_{adi}$ which is limited by the adiabatic condition,
and the time $\tau_{iso}$ allocated to the {\em isochores}. 

Optimizing the time allocation on the {\em isochore}s subject to (\ref{eq: d-tau optimum work}) leads to the optimal condition \cite{k221}:
\begin{equation}
\Gamma_{c}\tau_{cyc}(\cosh(\Gamma_{h}\tau_{h})-1)=
\sinh(\Gamma_{h}\tau_{h}+\Gamma_{c}\tau_{c})-\sinh(\Gamma_{c}\tau_{c})-\sinh(\Gamma_{h}\tau_{h})~.
\label{eq:quasistatic optimal tau}
\end{equation}
When $\Gamma_{h}=\Gamma_{c}\equiv\Gamma$ this expression simplifies to:
\begin{equation}
2x+\Gamma\tau_{adi}=2\sinh(x)
\label{eq:ftf}
\end{equation}
(where $x=\Gamma_{c}\tau_{c}=\Gamma_{h}\tau_{h}$). For small $x$ Eq. (\ref{eq:ftf})
can be solved leading to the optimal time allocation on the {\em isochores}: 
$\tau_c=\tau_h \approx \left(\Gamma \tau_{adi}/3\right)^{\frac{1}{3}}/\Gamma$. Taking into consideration
the restriction on the adiabatic condition this time can be estimated to be:
$\tau_c=\tau_h \propto \frac{1}{\Gamma} \left(\frac{\Gamma}{J} \right)^{\frac{1}{3}}$.

We can now expect two limits for the optimal cooling power the first when $\Gamma$ is sufficiently large
the cycle time $\tau_{cyc}$ will be dominated by the time on the {\rm adiabats} then for large $\omega_c$
\begin{equation}
{\cal P}_c(max)  \propto  \hbar J^2 e^{-\frac{\hbar J}{k_b T_c}}
\label{eq:maxpower1}
\end{equation}
When the heat transfer time dominates, $\tau_c > \tau_{hc} $ then:
\begin{equation}
{\cal P}_c(max)  \propto  \hbar \frac{J^{\frac{4}{3}}}{\Gamma^{\frac{2}{3}}} e^{-\frac{\hbar J}{k_b T_c}}
\label{eq:maxpower2}
\end{equation}

Noise on the {\em adiabats} modifies the optimal time allocation. Phase noise has its minimum for large
values of $l$, Cf.  Eq. (\ref{eq:phasnoise}). It approaches this minimum after a few revolutions independent of $K_{hc}$.
The optimum power is a compromise between large time 
allocation on the {\em adiabat} to reach minimize noise and small cycle time to maximize power. 
As a result the scaling $\tau_{adi} \propto 1/J $ is still maintained, therefore Eq. (\ref{eq:maxpower1}) or Eq. (\ref{eq:maxpower2}) will hold.
For amplitude noise the minimum $\delta$ is obtained for the minimum time frictionless solution which also leads to the scaling of power as in Eq. (\ref{eq:maxpower1}).

\section{Simulating the cycle}

After the segment propagators have been solved the cycle propagator can be assembled.
For constant $m$ the cycle propagator ${\cal U}_{cyc}$ has 
a closed form solution. Other scheduling functions $\omega(t)$ require numerical integration
of the equation of motion Eq. (\ref{eq:newanad1}). We have verified that our numerical integration coincides with the analytic expressions when available. 

The purpose of the simulation is to determine the optimal performance of the refrigerator.
The cooling power was extracted from the limit cycle obtained by propagating the cycle iteratively from an initial state until convergence.  The optimal cooling power was studied as a function of total cycle time $\tau$. 
For a fixed cycle time the heat extracted ${\cal P}_c$ was optimized with respect to the time allocation
on each segment. A random search procedure was used for this task.

In general two types of cycles emerge classified according to the cycle time.
The first are sudden cycles with very short periods $\tau_{cyc} \ll \frac{2 \pi}{\Omega}$ 
which are characterized by a global topology. 
These cycles are not presented and will be addressed separately \cite{tova09}.
The focus of the present study are cycles with a period comparable or longer
than the internal time scale $\tau_{cyc}  > 2 \pi / \Omega$. 
The cycles of optimal cooling rate and minimum temperature are of this type.

\begin{figure}[tb]
\vspace{-0.66cm}
\hspace{2.cm}
\center{\includegraphics[height=8cm]{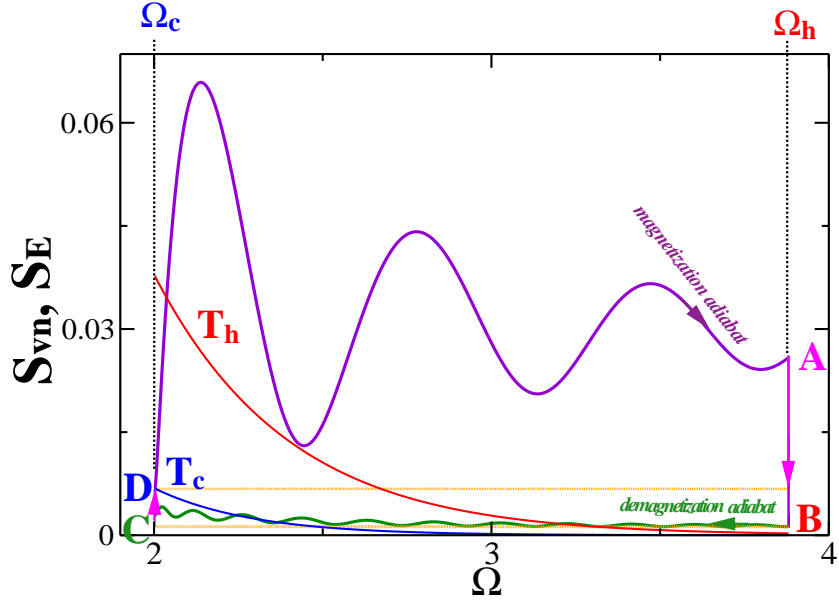}}
\vspace{0.3cm}  
\caption{Typical optimal cycle of refrigerator with linear scheduling,
with $T_c > T_c^{min} $, in the 1.($\Omega, S_E$),  
2. ($\Omega, S_{VN}$) plane (lower rectangle). The isotherms corresponding   
to the cold and hot baths temperatures, $T_c$ and $T_h$ are indicated.
The difference between the energy entropy and the von Neuman
entropy is the result of quantum friction: point A is higher than point D and point C is higher than point B.
The parameters are: The cycle parameters are:  $J=2$, $T_c=0.18 $, $T_h=0.24$, $\omega_c=0.1$, $\omega_h=3.325$, 
$\tau_c=12.292$, $\tau_h=11.615$, $\tau_{hc}=18.016$, $\tau_{ch}=5.077$. }
\label{fig:typical}  
\end{figure}

\begin{figure}[tb]
\vspace{-0.66cm}
\hspace{2.cm}
\center{\includegraphics[height=8cm]{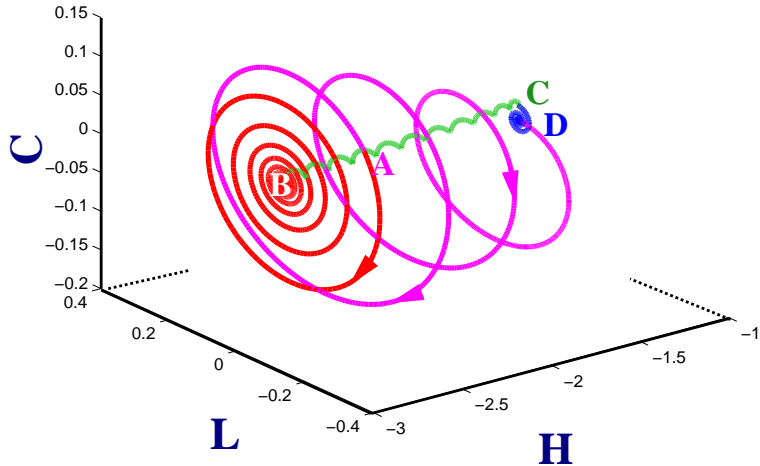}}
\vspace{0.3cm}  
\caption{Typical optimal cycle trajectory with linear scheduling
shown in the  $\Op H$, $\Op L$ and $\Op C$ coordinates,
for the same  parameters as Fig. \ref{fig:typical}. Point $A$ represents the beginning of the hot {\em isochore}.
Point $B$ represents the beginning of the demagnetization {\em adiabat}. Point $C$ represents the beginning of the cold {\em isochore}. Point $D$ represents the beginning of the magnetization {\em adiabat}.
Notice the big difference between the demagnetization and magnetization adiabats.}
\label{fig:traj}  
\end{figure}

Fig. \ref{fig:typical} and Fig. \ref{fig:traj} present  a typical cycle constructed with $\omega(t)$ 
linear in $t$ with optimal time allocation. 
Fig. \ref{fig:typical} displays the entropy frequency plane. Fig. \ref{fig:traj} shows the trajectory in the $\Op H$, $\Op L$ and $\Op C$ coordinates.
The positioning of the cycle with respect to the hot and cold isotherms
shows that it operates as a refrigerator with positive ${\cal Q}_c$.
The end point $D$ of the cold \em isochore \rm is below the
equilibrium point with the cold bath. On the scale of Fig. \ref{fig:typical} this is hard to observe.
One should also notice that
$S_{VN}$ which is constant on the \em adiabats \rm, and always a lower
bound to $S_E$, almost touches the minimal $S_E$ of the {\em adiabats}.
The vertical distance from D to A, and from B to C is the result of quantum friction.  

\begin{figure}[tb]
\vspace{-0.66cm}
\hspace{2.cm}
\center{\includegraphics[height=8cm]{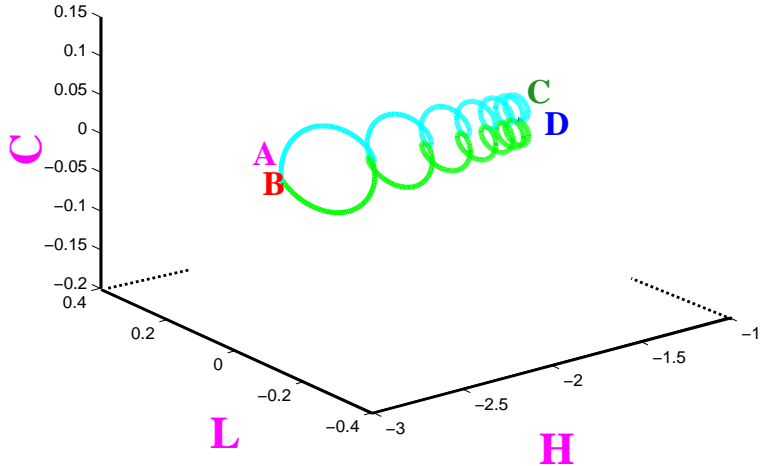}}
\vspace{0.3cm}  
\caption{Typical optimal cycle trajectory 
shown in the  $\Op H$, $\Op L$ and $\Op C$ coordinates, corresponding to Fig. \ref{fig:1}
Notice that on this scale the {\em isochores} are barely observable. }
\label{fig:traj2}  
\end{figure}

The asymmetry between the demagnetization and magnetization {\em adiabats} 
can be noticed in both figure \ref{fig:typical} and figure \ref{fig:traj}. The reason for this asymmetry 
is that the heat caused by friction in the magnetization {\em adiabat} can be dissipated
to the hot bath. This is not true  on the demagnetization {\em adiabat} where friction 
limits the possibility of heat extraction.
This leads to very different time allocation $\tau_{hc} > \tau_{ch}$.
The linear scheduling cycle should be compared to the cycle Fig. \ref{fig:1} and Fig \ref{fig:traj2} where the friction is limited
due to the quantization.
The obvious difference is the symmetry between the demagnetization and magnetization {\em adiabats}.
At the beginning and the end of the frictionless segments the von Neumann and the energy entropy coincide.
Periodic dynamics on the {\em adiabats} is also observed for optimal linear scheduling Cf. Fig \ref{fig:traj}, nevertheless frictionless solutions are not obtained.

\subsection{Numerical experiments}

\begin{figure}[tb]
\vspace{1.2cm} 
\center{\includegraphics[height=8cm]{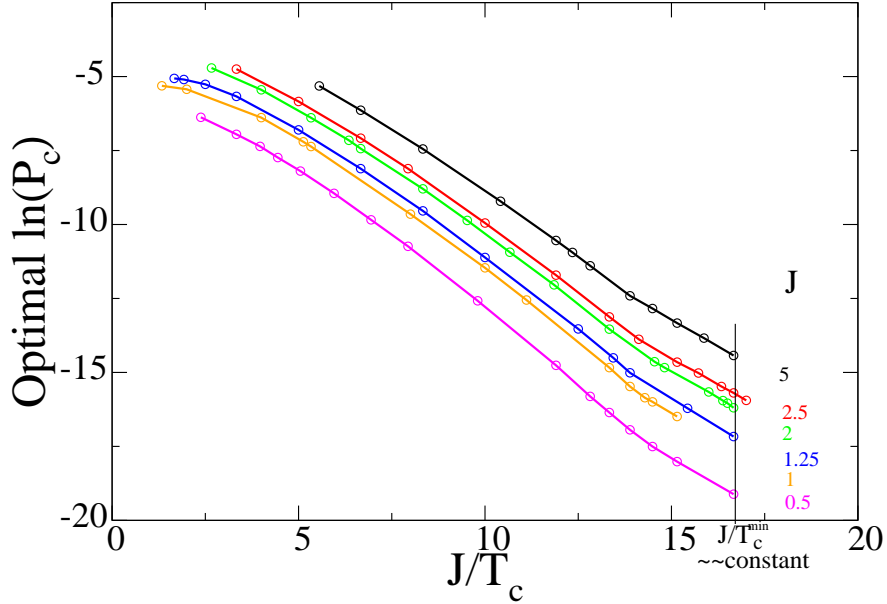}}
\caption{ 
The logarithm of the optimal heat power ${\cal P}_c$
as a function of $ J/T_c$  for different $J$ values (linear scheduling).
On the graphs the ratio ${\cal R}=(T_c \Omega_h)/(T_h \Omega_c)=1.453$ is kept constant
where $\omega_c=0.1$ and $T_c/T_h=0.75$. }
\label{fig:rT075p01}   
\end{figure}

We studied the optimal cooling cycles for a very large set of parameters 
for different scheduling functions.  
Fig \ref{fig:rT075p01} shows the optimal cooling power as a function of $J/T_c$ where the ratio
${\cal R}=\frac{T_c \Omega_h}{T_h \Omega_c} $ was maintained constant. The parameter ${\cal R}$  
addresses the distance of the operation conditions from the reversible limit where ${\cal R}=1$. 
The simulations were performed for a  predefined ${\cal R} > 1$ so that the second law is never violated
Cf. Eq. (\ref{eq:basicineq}).
Fig. \ref{fig:rT075p01} was obtained for a linear scheduling function of $\omega(t)$ without the  addition of noise. 
The sticking feature is that all graphs for different $J$ values terminate at the same minimum $J/T_c$.
This graph has initiated our search for a possible explanation. In retrospect it represents the influence of uncontrolled numerical noise. Comparing to Eq. (\ref{eq:mintemp}) we can estimate the value of $\delta$ as $\approx 10^{-7}$. In addition all lines  corresponding to different $J$ values can be collapsed by shifting
vertically by $\log J^2$. This finding  
shows consistency with the scaling of ${\cal P}_c$ with $J^2$ Cf. Eq. (\ref{eq:maxpower1}).

\begin{figure}[tb]
\vspace{-0.66cm}
\hspace{2.cm}
\center{\includegraphics[height=8cm]{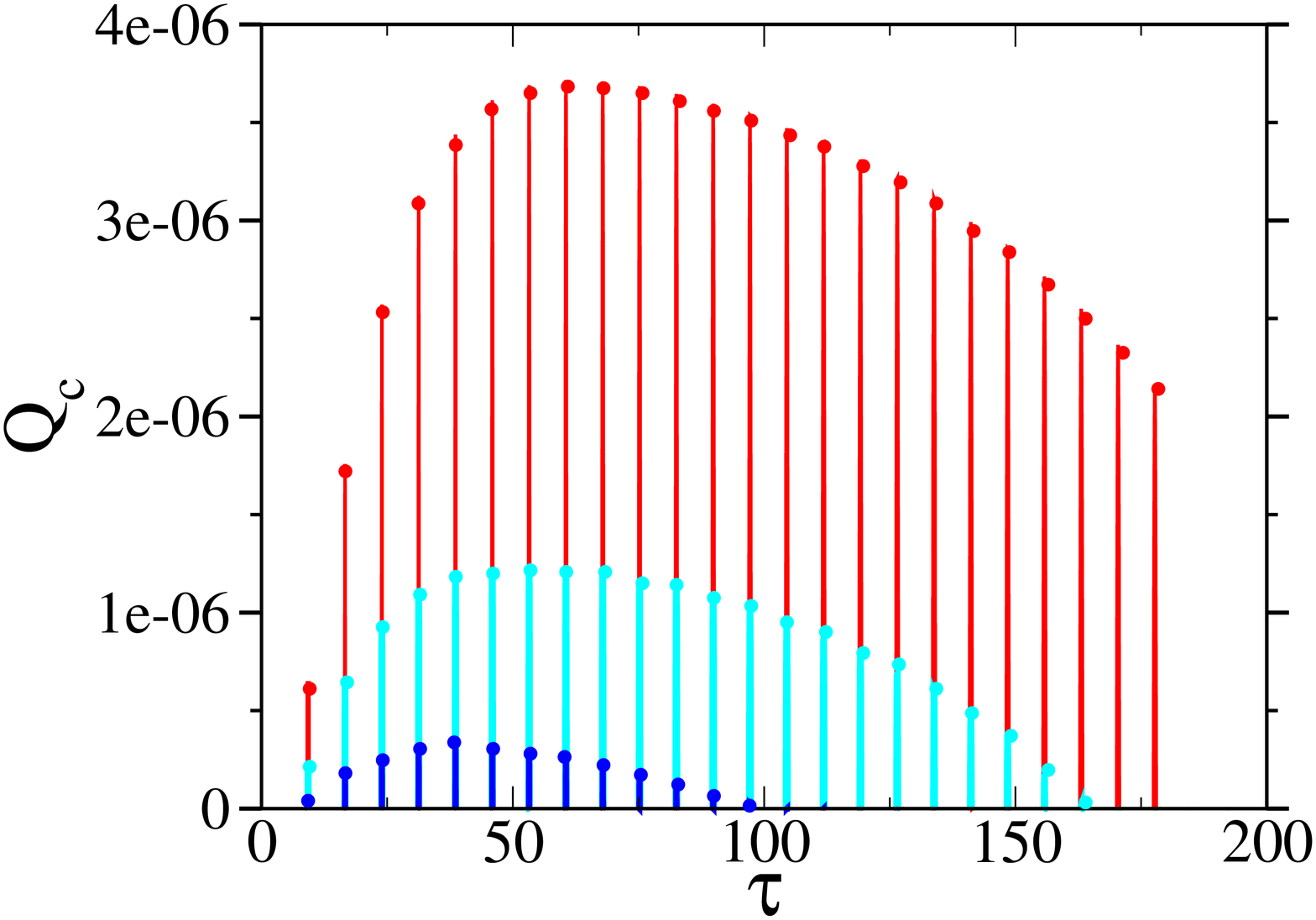}}
\vspace{0.3cm}  
\caption{The optimal  heat extracted ${\cal Q}_C$ as a function of cycle time for three sets of temperatures:
$T_c= 0.105$, $T_h=0.14$ (Top: red)
$T_c= 0.0975$, $T_h=0.13$(Middle: purple)
$T_c= 0.09$, $T_h=0.12$(Bottom: magenta).  Results obtained by random search for stationary $m$
with the restriction of $\tau_{hc}=\tau_{ch}$. Other parameters $j=2$ $\omega_c=0.1$, $\omega_h=3.32576$ and
no added noise.}
\label{fig:optq}  
\end{figure}

Simulations with constant $m$ confirm the quantizations behavior of the optimal conditions.
Figure \ref{fig:optq} displays ${\cal Q}_c$ for optimal cycles as a function of total 
cycle time $\tau$. The quantization of the cycle time is apparent corresponding to almost
frictionless complete revolutions on the {\em adiabats}.
The comb like function ${\cal Q}_c^{opt}(\tau)$ has a maximum at approximately $l=8$ at the high temperature
and $l=6$ at the lowest temperature which is very close to the minimum temperature possible in these simulations. 
The maximum of ${\cal Q}_c^{opt}(\tau)$ is an indication of uncontrolled numerical noise in the simulation.
Noiseless operation conditions would result in a flat comb distribution of ${\cal Q}_c^{opt}(\tau)$.
\begin{figure}[tb]
\vspace{-0.66cm}
\hspace{2.cm}
\center{\includegraphics[height=8cm]{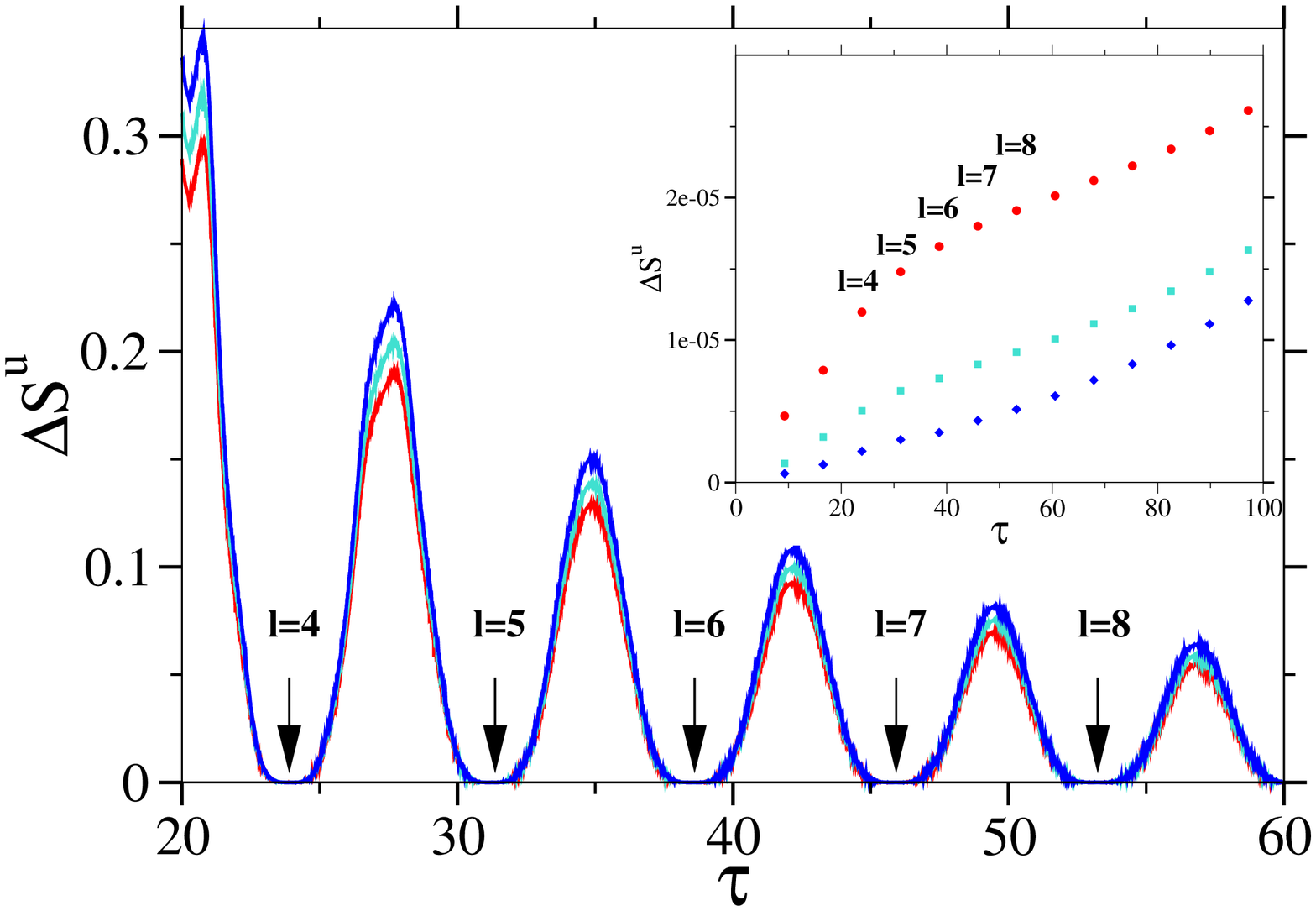}}
\vspace{0.3cm}  
\caption{The entropy production ${\cal S}^u$ as a function of cycle time for three sets of temperatures
corresponding to Fig. \ref{fig:optq}. Only the cycles with almost zero ${\cal S}^u$ function as refrigerators.
Their entropy production is shown in the insert on a very different scale. The quantum numbers are indicated.
The break in the slope corresponds to the maximum ${\cal P}_c$ in  Fig. \ref{fig:optq}.
}
\label{fig:dsu}  
\end{figure}

Fig. \ref{fig:dsu} shows the entropy production as a function of cycle time. Only the cycles with very small entropy production
corresponding to almost frictionless cycles, operate as refrigerators. All non quantized cycles have a very large entropy production
which decreases when the cycle time becomes longer. 
The quantized cycle (insert) have a very small entropy production which increases with the cycle time $\tau$. 
Fig. \ref{fig:dsu} demonstrates that quantum friction is accompanied by large entropy production.

We attempted  to identify the character of the numerical noise in the simulation. 
The procedure was to estimate the minimum temperature for a set of parameters $J$ ${\cal R}$  and 
${\cal C}= \Omega_c/\Omega_h$ the magnetization ratio. Then we used Eq. (\ref{eq:maxq}) to estimate $\delta$.
From the functional dependence of $\delta$  on the parameters we tried to empirically asses the 
numerical noise in the simulation. In general we found both phase and amplitude noise. This can be observed
in the trimming of both the high and low $l$ ends of the comb in Fig. \ref{fig:optq}. 
In general we found that $\delta$ decreased with ${\cal C}$ the compression ratio and with ${\cal R}$
the deviation from reversibility.  These dependencies were found for both constant 
$m$ and linear scheduling where the constant $m$ resulted consistently with a lower minimum temperature.
The findings indicate that there is an additional 
source of numerical noise beyond the amplitude and phase noise.  

\begin{figure}[tb]
\vspace{-0.66cm}
\hspace{2.cm}
\center{\includegraphics[height=8cm]{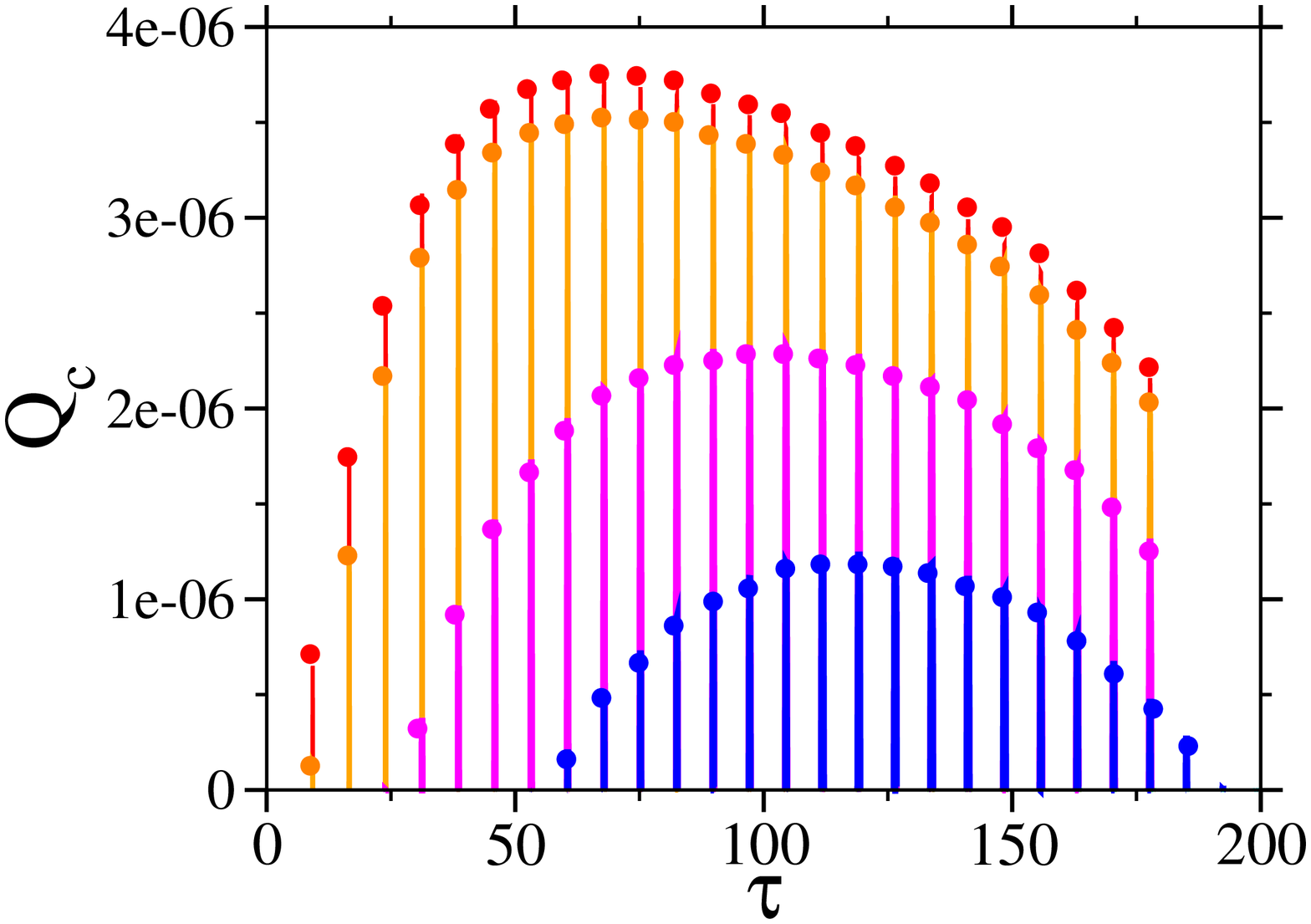}}
\vspace{0.3cm}  
\caption{The optimal  heat extracted ${\cal Q}_C$ as a function of cycle time for three values of external phase  noise $\gamma_p$:
Top red $\gamma_p=0$,  middle orange $\gamma_p=10^{-6}$, 
middle magenta $\gamma_p=10^{-5}$, bottom blue $\gamma_p= 2 \cdot 10^{-5}$. $\gamma_p=5 \cdot 10^{-5}$ did not result in positive ${\cal Q}_c$. Other parameters are as in Fig. \ref{fig:optq}.}
\label{fig:optqn}  
\end{figure}

The existence of uncontrolled numerical noise hinders the study of the 
additional effects of the imposed phase and amplitude noise.
The cycle simulation was repeated with the addition of phase noise Cf. Eq. (\ref{eq:newanad}). 
As can be seen in Fig. \ref{fig:optqn} an increasing amount of phase noise depresses ${\cal Q}_c$ and moves the maximum to larger $l$, Cf. Eq. (\ref{eq:sivuv}). Numerical noise trims the high values of $l$.

\begin{figure}[tb]
\vspace{-0.66cm}
\hspace{2.cm}
\center{\includegraphics[height=8cm]{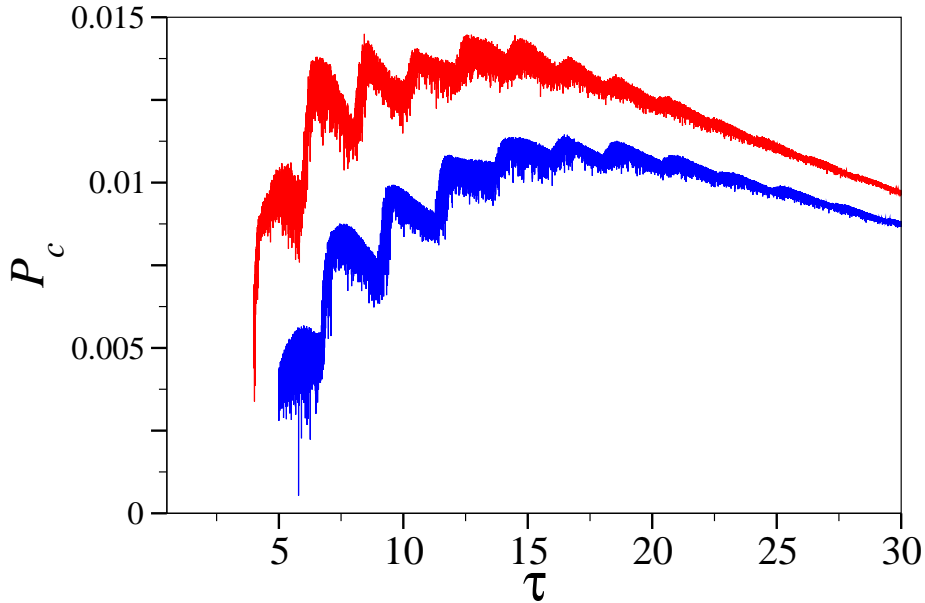}}
\vspace{0.3cm}  
\caption{The optimal cooling power  ${\cal P}_C$ as a function of cycle time for linear scheduling. 
Upper plot (red) $\omega_c=0.5$ and $\omega_h=3.9$. Lower plot (blue)  $\omega_c=0.1$, $\omega_h=3.4$.
$J=1.25$ $T_c=0.725$, $T_h=0.966$.}
\label{fig:optpower}  
\end{figure}

The quantization of the optimal cycles is independent of the specific scheduling. 
When the cold bath temperature $T_c$ is increased the quantization of ${\cal Q}_c$ and ${\cal P}_c$ is less
pronounced. This can be seen in Fig \ref{fig:optpower} where the optimal power is plotted as a function of cycle time $\tau$ for linear scheduling. The sharp comb structures of Fig. \ref{fig:optq} is replaced by periodic modulation on top of a continuous background. At higher temperatures cycle with more friction can still operate as refrigerators and the quantization features are washed out.

\section{Summary}
\label{sec:conclus}

Quantum friction is the result of the inability of the system to follow adiabatically the time dependent 
changes in the control Hamiltonian $[\Op H(t),\Op H(t')] \ne 0$. As a result the state of the system will develop non diagonal terms in the 
energy representation $\Op \rho_e$. The signature of this phenomena is an increase in the energy entropy ${\cal S}_E$.
The key to cold temperature refrigeration is no increase in energy in the demagnetization segment beyond the adiabatic limit.
Perfect adiabatic following which requires infinite time will lead to frictionless demagnetization. 
Under conditions that the second law is fulfilled  ${\cal R} > 1$ the cooling can continue to $T_c =0$. 
We then introduced an adiabatic measure $\mu$ to characterize the instantaneous nonadiabatic transition rate.
The limit $\mu \rightarrow 0$ corresponds to perfect adiabatic following.  Our first surprise was that 
constant $\mu$ led to closed form solutions for the dynamics. Moreover these solutions unraveled a
quantized family of frictionless solutions $\delta=0$. These solutions are characterized by a state 
$\Op \rho$ commuting with the Hamiltonian  at the beginning and end of the segment 
$[\Op H(0), \Op \rho]=[\Op H(\tau_{hc}), \Op \rho]=0$. 
Similar frictionless solutions were found for a working medium constructed from harmonic oscillators \cite{k242,k243}.
These frictionless solution can be carried out in a finite cycle time i.e.  the cooling power does not vanish ${\cal P}_c > 0$. 
If these frictionless cycles could be realized they could operate to $T_c=0$.

When we tried to simulate numerically  the frictionless cycles we got into conflict. Any attempt resulted in 
a minimum temperature $T_c(min)$ which scaled linearly with  the energy gap $\hbar J$. This observation 
eventually led us to the realization that any cycle is subject to noise. To follow this idea we constructed a model for 
external noise on the controls. Amplitude noise is the result of fluctuations in the  magnitude of the external magnetic field.
Since this noise term does not commute with the Hamiltonian it is not surprising that it will destroy the adiabticity,
 leading to $\delta > 0$. The surprise was the devastating effect of phase noise which commutes with 
 the instantaneous Hamiltonian. Such a term can be the result of weak continuous measurement of energy
 during the {\em adiabats}. Such a measurement leads to partial collapse of the state to the energy representation.
 Naively one would expect this to lead to frictionless solutions. We have employed such an idea successfully to reduce
 friction in a quantum engine \cite{k215}. Nevertheless for a refrigeration cycle close to its minimum temperature 
 phase noise accumulates leading to $\delta > 0$.  Both types of noise are sufficient to eliminate frictionless solutions 
 including the perfect  infinite time adiabatic following frictionless cycle.
 
\begin{figure}[tb]
\vspace{1.2cm} 
\center{\includegraphics[height=7cm]{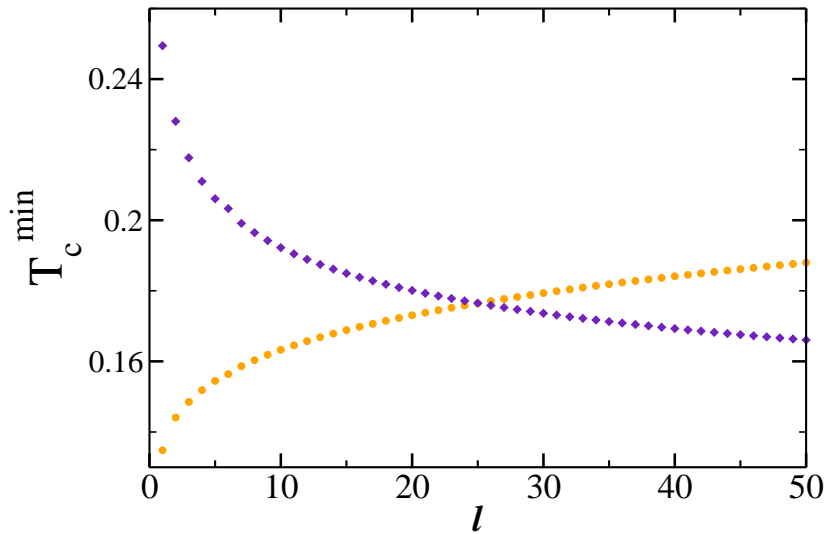}}
\caption{ 
The minimum temperature as a function of the quantization number $l$. The diamonds represent phase noise
and the circles amplitude noise. $J=2$
 }
\label{fig:mintemp}   
\end{figure}

Once the devastating effect of noise is appreciated it can be directly linked to a restriction on
the minimum temperature.
The minimum temperature $T_c^{min}$ depends on $-1/\log \delta$ Cf. Eq. (\ref{eq:mintemp})
and will be on the order of the energy gap $\hbar J$.
This finding is consistent with experiments on demagnetization cooling of a gas \cite{pfau06}
which obtained a minimum temperature an order of magnitude larger than the theoretical prediction \cite{pfau05} attributing the discrepancy to noise in the controls. 
Figure \ref{fig:mintemp} shows the dependence of the minimum temperature of a refrigerator subject to phase and amplitude noise.
The minimum temperature is related  to the quantum number of the frictionless solutions $l$. 
The two types of noise show an opposite dependence on $l$.
Amplitude noise favors small cycle times $l=1$ while phase noise favors small $m$ meaning large $l$.
If both types of noise are operative the minimum temperature will be a compromise in Fig. \ref{fig:mintemp}
at $l=25$.  Other sources of noise will also limit $T_c(min)$, for example our study was hindered by numerical noise.

To conclude, it seems that any refrigerator constructed with a working 
medium possessing an uncontrolled energy gap will reach a minimum operating temperature
on the order of the energy gap.  For a working medium that has a controllable gap we found that if the gap 
is linear with  $T_c$ there is no minimum temperature above $T_c=0$ if the gap can be reduced to zero \cite{k243}.
\pagebreak
\section*{Aknowledgements}
We want to thank Yair Rezek, Peter Salamon and Lajos Diosi for crucial discussions.
This work is supported by the Israel Science Foundation.

\appendix
\section{Optimality of constant $\mu$}
\label{app:A}

We show that constant $\mu$ is the minimum of the non adiabatic deviations i.e. minimum of of $\delta$.
We can transform Eq. (\ref{eq:m}) to  the differential equality: $\mu dt = \frac{d \omega}{\Omega^3}$ leading to:
\begin{equation}
\int_0^{\tau_{hc}} \mu(t) dt ~~=~~ \int_{\omega_h}^{\omega_c} 
\frac{d \omega}{\Omega^3}
\label{eq:mu1}
\end{equation}
We decompose $\mu$ to a constant and a time dependent part $\mu=\mu_0 +\mu_1 g(t) $.  
Without loss of generality we impose $\mu_0 \tau_{hc} =\int_{\omega_h}^{\omega_c} 
\frac{d \omega}{\Omega^3}$, then $\int_{0}^{\tau_{hc}} g(t) dt=0$. 

The first order correction to the propagator ${\cal U}_{2}$ due to time depdentce in $\mu$ is the time average 
$(1/\tau_{hc}) \int_0^{\tau_{hc}} dt {\cal U}_2(t)$. This will translate to a time average of $\delta$. 
The dependence of  $\delta$ Eq. (\ref{eq:delta})  and Eq. (\ref{eq:calprop}) on $\mu$ is:
\begin{equation}
\delta = \mu^2 \frac{(1-c)}{1+\mu^2}~~,
\label{eq:delta2}
\end{equation}
which is a monotonic increasing function of  $\mu^2$ with minimum at $\mu^2=0$. 
The first order correction to ${\cal U}_2$ will lead to $\delta = \delta_0 +\delta_1$ where $\delta_0$ 
is the stationary result. Then  expanding in $\mu_1$ will lead to:
$\delta_1 = \left(\frac{1-c}{1+\mu_0^2} \right)~\frac{\mu_1^2}{\tau_{hc}} \int_0^{\tau_{hc}}  g^2(t) d t $
which is positive definite, therefore a stationary $\mu$ is a minimum of $\delta$.

\bibliography{../dephc1,../../../Database/pub}

\end{document}